\documentclass[8pt]{article}
\usepackage{graphicx}
\usepackage{pgfplots}
\usepackage{newlfont}
\usepackage[colorlinks,linkcolor=blue,citecolor=blue,anchorcolor
=blue,bookmarksopen,pdfpagetransition={Wipe}]{hyperref}
\usepackage{amsmath,amsthm,amssymb}
\usepackage{subfigure}
\usepackage{amsmath}
\usepackage[linesnumbered,ruled]{algorithm2e}
\newtheorem{thm}{Theorem}[section]

\newtheorem{defn}[thm]{Definition}

\numberwithin{equation}{section}
\begin{document}
\title{\textbf{Multi-color image compression-encryption algorithm based on chaotic system and  fuzzy transform }}
\author{M. Zarebnia$^a$\footnote{zarebnia@uma.ac.ir}\
\, R. Kianfar$^a$\footnote{rahele.kianfar@student.uma.ac.ir}\
, R. Parvaz$^a$\footnote{Corresponding author: rparvaz@uma.ac.ir }
}
\date{}
\maketitle
\begin{center}
$^a$Department of Mathematics, University of Mohaghegh Ardabili,
56199-11367 Ardabil, Iran.\\
\end{center}
\begin{abstract}
In this paper an algorithm for multi-color image compression-encryption is introduced.
 For compression step fuzzy transform based on exponential b-spline  function is used.
In encryption step, a novel combination chaotic system based on Sine and Tent systems is proposed.
Also in the encryption algorithm, 3D shift based on chaotic system is introduced. 
The simulation results and security analysis show that the proposed algorithm is secure and efficient.

\indent
\end{abstract}
\vskip .3cm

\section{Introduction}
\indent \hskip .65cm
In recent years with the development of the information transmission,
fast and secure transmission have become important subject.
Various type of methods  for compression-encryption and encryption  have been studied in  \cite{1,2,3,4,5,6}.
Also  multi- image encryption  have been studied in \cite{7,8}.
In this work, in the first step we use the fuzzy transform for image compression step.
Lossy image compression and reconstruction basis on  fuzzy transform has been proposed in \cite{9,10,11}.
For fuzzy  partition in fuzzy transform method, exponential b-spline function is used,
more details about this function can be found in
\cite{12,14}.
 In the next step, for encryption algorithm,
a combination chaotic system is introduced.
This system is introduced in \cite{15}.  In this system for all values of $r\in (0,4]$ the Lyapunov exponent is positive.
Also the combination chaotic system have uniform distribution over output range.
In the encryption algorithm, by using  chaotic system, we define two-dimensional block matrix shift and 
three-dimensional matrix shift.
Then by using these matrices, images are scrambled.\\
 
The organization of this paper is as follows: In Section 2, image compression and reconstruction by using fuzzy transform
is explained. In Section 3, we introduce the chaotic system and
the color image encryption algorithm. Experimental results and algorithm analysis
are given in Section 4. A summary is given at the end of the paper in Section 5.
\section{Fuzzy transform based on exponential  b-spline}
In this section, we describe fuzzy transform based on exponential b-spline.
The basis of exponential b-spline has been proposed in \cite{12,14}. 
For non-decreasing sequence of knots as $x_1\leq x_2\leq\cdots \leq x_n$, exponential B-splines of order $2$ are defined as follows

\begin{align}
B^2_{0}(x):=
\left\{%
\begin{array}{ll}
\frac{sinh(\rho_{1}(x_{2}-x))}{sinh(\rho_{1} h_{1})},&x\in[x_{1},x_{2}],\\
\\
0,&~otherwise,\\
\end{array}%
\right.
\end{align}
\begin{align}
B^2_{n}(x):=
\left\{%
\begin{array}{ll}
\frac{sinh(\rho_{n-1} (x-x_{n-1}))}{sinh(\rho_{n-1} h_{n-1})},&x\in[x_{n-1},x_{n}],\\
\\
0,&~otherwise.\\
\end{array}%
\right.
\end{align}
Also for $i=1,\cdots,n-1$, we define
\begin{align}
B^2_{i}(x):=
\left\{%
\begin{array}{ll}
\frac{sinh(\rho_{i-1}(x-x_{i-1}))}{sinh(\rho_{i-1} h_{i-1})},&x\in[x_{i-1},x_{i}],\\
\\
\frac{sinh(\rho_{i}(x_{i+1}-x))}{sinh(\rho_{i} h_{i})},&x\in[x_{i},x_{i+1}],\\
\\
0,&~otherwise,\\
\end{array}%
\right.
\end{align}

where $\rho_i~(i=1,\ldots,n-1)$ are tension parameters and $h_i:=x_{i+1}-x_{i}$.
The tension parameter in b-spline function plays a important role in the slope of the function (see Fig.1).
Also  exponential B-splines of 
arbitrary
order  can be found as follows
\begin{align*}
B^k_j(x):=\frac{\int^x_{x_j}B^{k-1}_j(y) dy}
{\sigma^{k-1}_j}-\frac{\int^x_{x_{j+1}}B^{k-1}_{j+1}(y) dy}{\sigma^{k-1}_{j+1}},~k=3,4,\ldots,
\end{align*}
where
\begin{align*}
\sigma^{k-1}_{j}:=\int^{x_{j+k-1}}_{x_j}B^{k-1}_{j}(y)dy.
\end{align*}

\begin{center}
\begin{figure}
\centering
\includegraphics[]{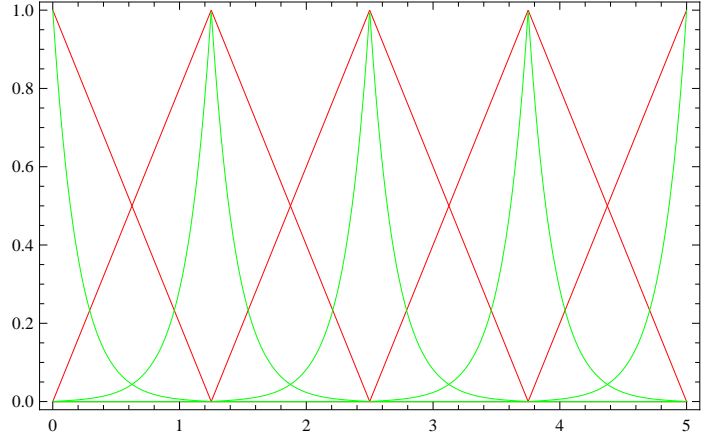}
\emph{\caption{Comparisons between exponential
 b-spline for different values of tension parameters, red lines for $\rho_i=0.001$ and  green lines for $\rho_i=5,~(i=1,\cdots,5)$.  }}
\end{figure}
\end{center}
The first study in fuzzy transform is introduced by Irina Perfilieva \cite{9}. According to \cite{9},  Fuzzy partition and F-transforms (Fuzzy transforms) are defined as follows.
\begin{defn} \label{def1}
\cite{9} Let $a=x_1<\cdots<x_n=b, (n\geq 2)$ be fixed nodes within $[a,b]$. We say that fuzzy sets $A_1,\cdots,A_n$, identified with their memberhip functions $A_1(x),\cdots,A_n(x)$, defined on $[a,b]$, form a fuzzy partition of $[a,b]$, if the fulfill the following conditions for
$k=1,\cdots,n$:\\

1-$A_k [a,b]\longrightarrow [0,1], A_k(x_k)=1;$\\

2-$A_k(x)=0$ if $x\notin(x_{k-1},x_{k+1});$\\ 

3-$A_k$ is continuous;\\

4-$A_k,k=2,\cdots,n$, strictly increases on $[x_{k-1},x_{k}]$ and $A_k,k=1,\cdots,n-1$,
 strictly decreases on $[x_{k},x_{k+1}]$\\
 
5-for all $x\in[a,b]$, $\sum^{n}_{k=1}A_k(x)=1$.\\  

For a function $f$ be given at nodes $(p_i,q_i)\in [a,b]\times[c,d],~i=1,\cdots,N,~j=1,\cdots,M$, discrete F-transform is defined as 
\begin{align}
&F_{kl}=\frac{\sum_{j=1}^{M}
\sum_{i=1}^{N}f_l(p_i,q_j)A_k(p_i)C_l(q_j)}{\sum_{j=1}^{M}\sum_{i=1}^{N}A_k(p_i)C_l(q_j)},
~k=1,\ldots,n,~l=1,\ldots,m,   
\end{align}
where $A_1,\cdots A_n, C_1,\cdots,C_m$( $n<N, m<M$) are fuzzy partitions of $[a,b]$ and 
$[c,d]$, respectively.
Also the inverse of F-transform is defined as 
\begin{align}
f^{F}_{nm}(x,y)=\sum^n_{k=1}\sum^m_{l=1}F_{kl}A_k(x)C_l(y).   
\end{align}
\end{defn}
 
We can easily prove that $\{B^2_i\}^n_i$  is satisfied in the Definition \ref{def1}. In compression step
we consider $B^2_i$ as $A_i$ and $C_i$, then we use fuzzy transform for image compression.
\section{Construction of the algorithm}  
\subsection{Compression-encryption algorithm}
\begin{figure}
\centering
\subfigure[{}]{
\includegraphics*[width=.45\textwidth]{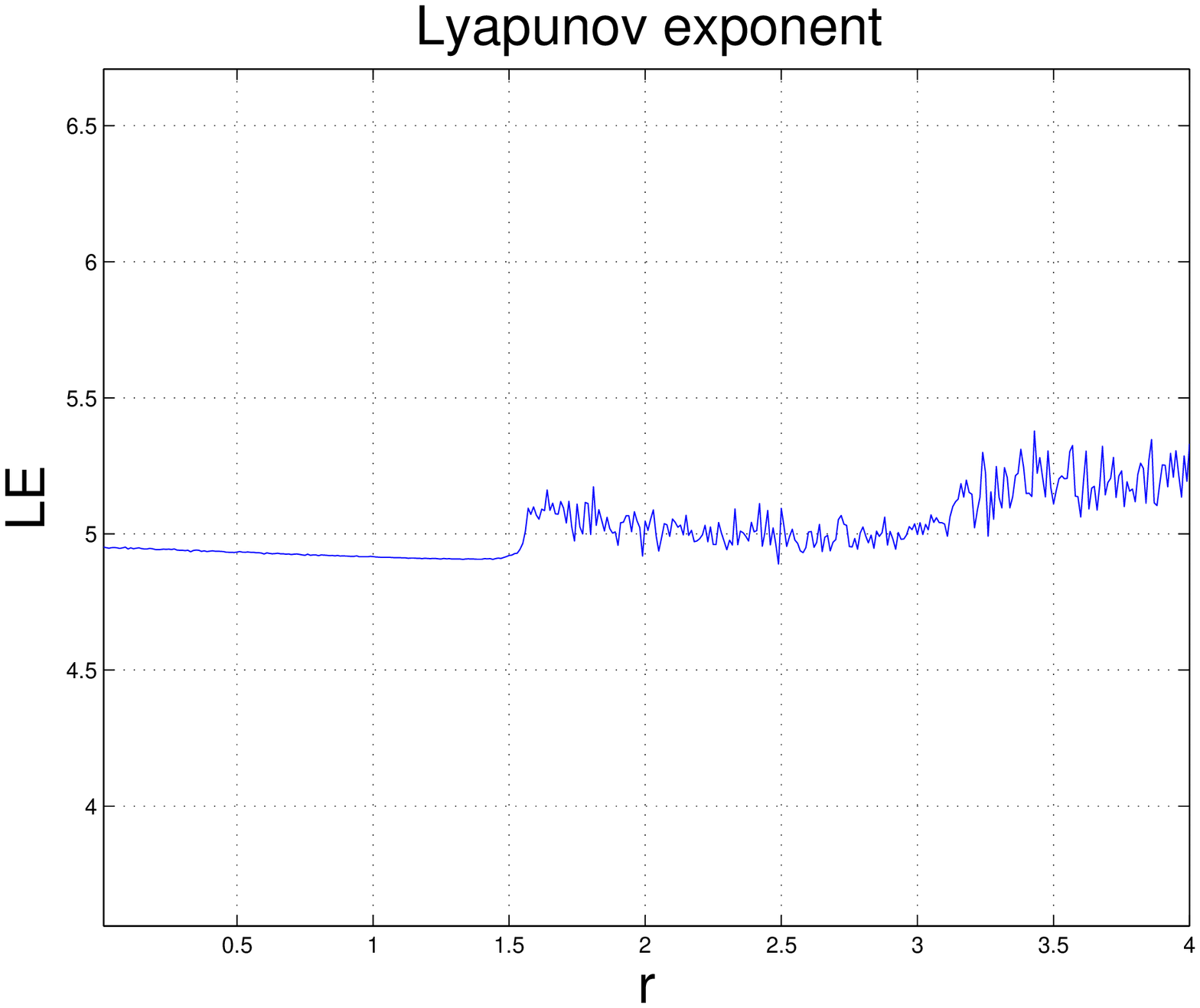}}
\hspace{2mm}
\subfigure[{}]{
\includegraphics*[width=.45\textwidth]{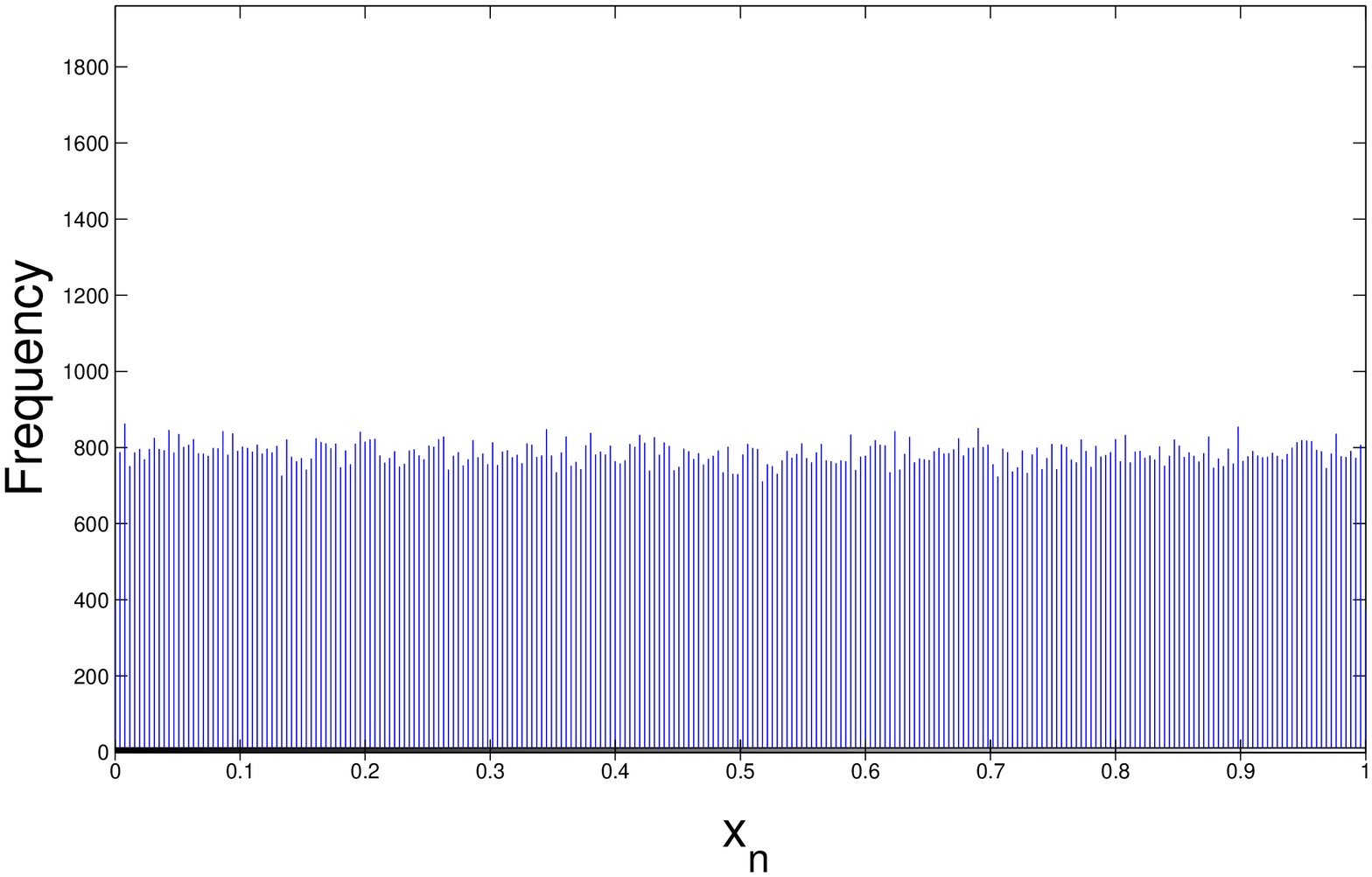}}
\subfigure[{}]{
\hspace{2mm}
\includegraphics*[width=.45\textwidth]{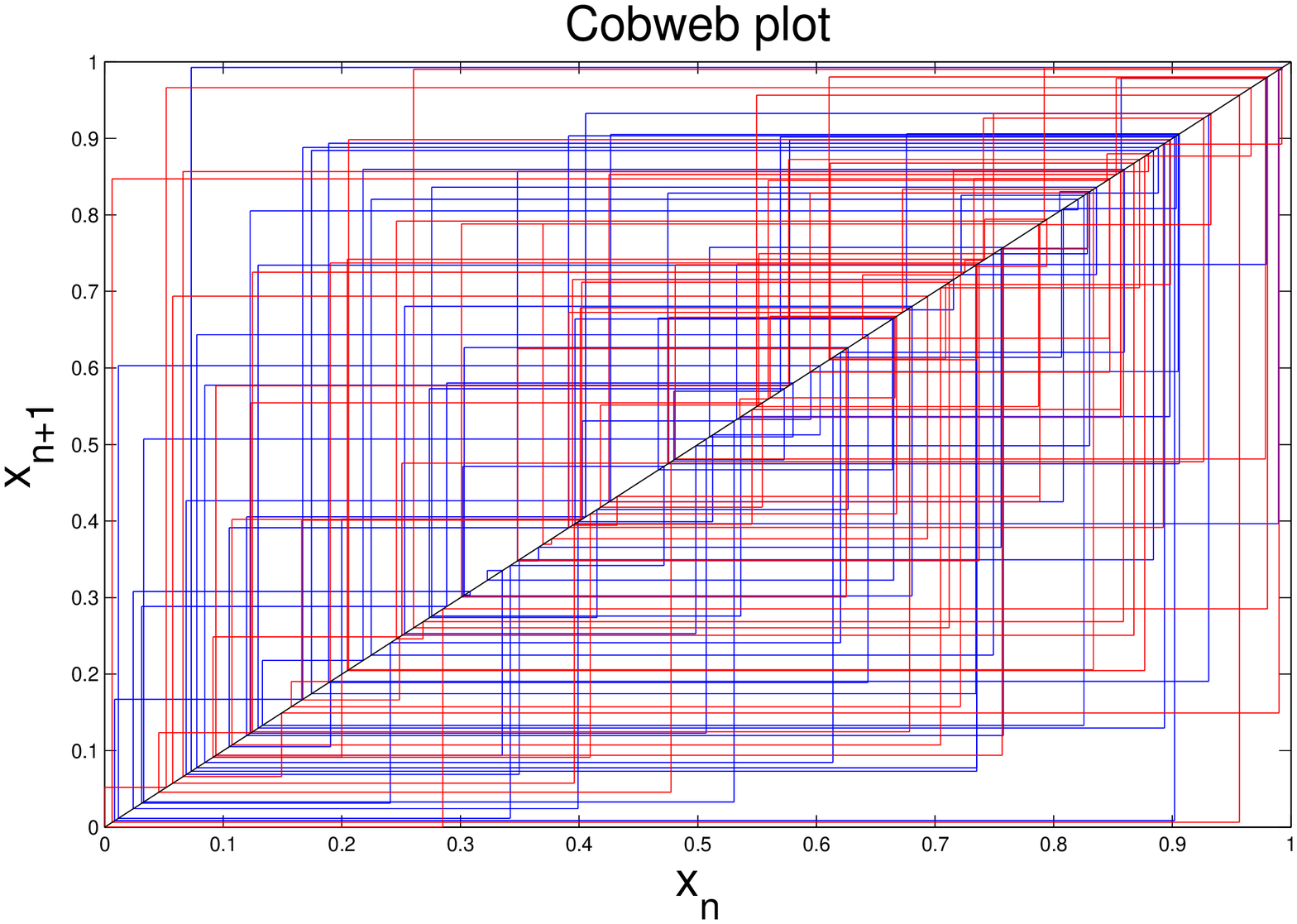}}
\hspace{2mm}
\subfigure[{}]{
\includegraphics*[width=.45\textwidth]{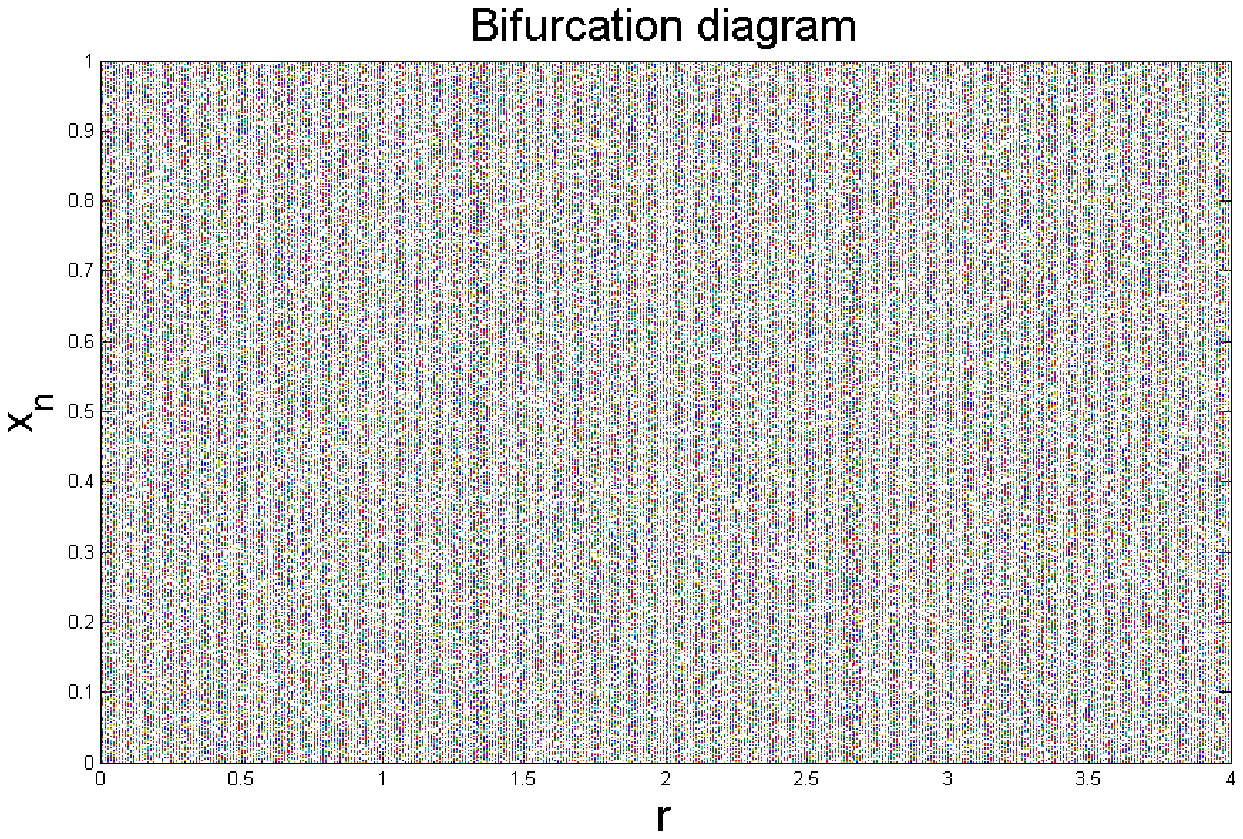}}
\emph{\caption{
 (a) Lyapunov exponent plot, (b) Histogram plot, (c) Cobweb plot,
(d) Bifurcation diagram.
}}\label{f-2}
\end{figure}

It is known that the histogram of the Logistic Tent system is
not flat enough.  For solving this problem we add weights and functions
in the Logistic Tent system or the Sine Tent system. For details, the reader can see \cite{15}. In this paper we consider 
a combination chaotic system based on Sine and Tent systems as follows
\begin{align}\label{e1}
X_{n+1}:=
\left\{%
\begin{array}{ll}
\omega_{1}f_{1}\circ F(r,X_n)+\alpha_{1}g_{1}(rX_n)+\xi_{1}\frac{(\beta_{1}-r)X_n}{2}~mod~1,~when~X_n<0.5,\\
\\
\omega_{2}f_{2}\circ F(r,X_n)+\alpha_{2}g_{2}(rX_n)+\xi_{2}\frac{(\beta_{2}-r)(1-X_n)}{2}~mod~1,~when~X_n\geq 0.5,\\
\end{array}%
\right.
\end{align}
where  $\omega_{1}=\omega_{2}=1,\alpha_{1}=\alpha_{2}=1,\xi_{1}=7,\xi_{2}=15,$  $\beta_{1}=40, \beta_{2}=20, f_{1}=\cos, f_{2}=\tan, g_{1}=\tan, g_{2}=x,F(r,x_n)=r\sin(\pi x_n)/4$.
The Lyapunov exponent,  Bifurcation, Histogram and Cobweb plots of combination chaotic system
are given in figure \ref{f-2}. From this figure we can see that for all values of $r$,
the Lyapunov exponent is positive.
The Cobweb plots  show chaotic behavior.
Also this figure show that the chaotic system has uniform distribution over output range.
The following definitions are used in proposed algorithm.

\begin{defn}
For an arbitrary vector $w\in R^{n m}$, matrix $\Delta(w)=(\delta_{i,j})\in R^{n\times m}$ is defined as follows
\begin{align*}
\delta(i,j):=w(i+(j-1)n),~i=1,\cdots,n,~j=1,\cdots,m.
\end{align*}
\end{defn}
\begin{defn}
For an arbitrary vector $w\in R^{3 n m }$, matrix $\Lambda(w)=(\lambda_{i,j,l})\in R^{n\times m\times 3}$ is defined as follows
\begin{align*}
\lambda(i,j,l):=w(i+(j-1)n+(l-1)nm),~i=1,\cdots,n,~j=1,\cdots,m,~l=1,2,3.
\end{align*}
\end{defn}
\begin{defn}
Consider matrix $A=(a_{i,j})\in R^{n\times m}$ such that
for all $(i,j)$ and $(k,l)$ with $(i,j)\neq(k,l)$, we have
$a_{i,j}\neq a_{k,l}$.
Suppose matrix $B =(b_{i,j})\in R^{n\times m}$ that for all $(i,j)$, $b_{i,j}=a_{k,l}$ for some  $(k,l)$.
For a block matrix $C=(C_{i,j})$ we define a block matrix
$\Upsilon^2_{A\rightarrow B}(C)=\big(D_{i,j}\big)$, where
$D_{i,j}=C_{k,l}$. 
\end{defn}
\begin{defn}
  Consider matrix $A=(a_{i,j,l})\in R^{n\times m \times 3}$ such that
for all $(i,j,d)$ and $(k,l,s)$ with $(i,j,d)\neq(k,l,s)$, we have
$a_{i,j,d}\neq a_{k,l,s}$.
Suppose matrix $B =(b_{i,j,d})\in R^{n\times m\times 3}$ that for all $(i,j,d)$, $b_{i,j,d}=a_{k,l,s}$ for some  $(k,l,s)$.
For a matrix $C=(c_{i,j,k})$ we define a matrix
$\Upsilon^2_{A\rightarrow B}(C)=\big(d_{i,j,k}\big)$, where
$d_{i,j,k}=	c_{k,l,s}$. 
\end{defn}
For above definitions we can say that
\begin{align*}
\Upsilon^l_{	B\rightarrow A}
\big(\Upsilon^l_{	A\rightarrow B}(C)\big)=\Upsilon^l_{	A\rightarrow B}\big(\Upsilon^l_{B\rightarrow A}(C)\big)=C,~~l=2,3.
\end{align*}
 
In this section we assume that we have a set of N images  as $I^1,\cdots,I^N$. Also let
the size of all the images are  $n\times m\times3$.
In the following algorithm $\mu(x_0,r,n)$ represents a vector as 
$(x_1,\cdots,x_{n})$
 where $x_i (i=1,\cdots,n)$ are 
defined by using (\ref{e1}) with $x_0$.
The compression-encryption algorithm
is written in the following steps.\\
\begin{center}
\begin{figure}
\centering
\includegraphics[width=110mm,scale=0.5]{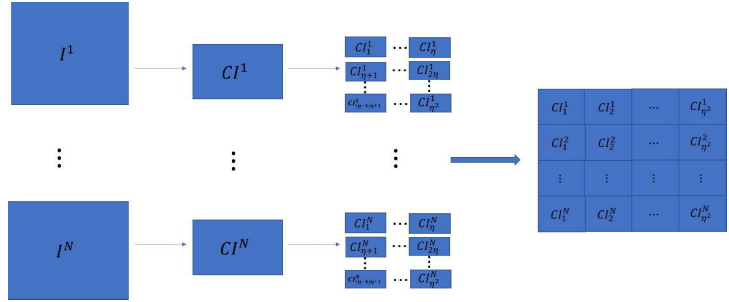}\label{f-3}
\emph{\caption{The step 1 and step 2 process of the proposed algorithm.  }}
\end{figure}
\end{center}     

\textbf{Step~1.} In the first step input images are converted into R(red), G(green), B(blue) component matrix. Then we 
use  exponential  b-spline fuzzy transform for each component part. We assume that the size of compressed image is
$nc\times mc\times 3$. In this step compressed images are considered as $CI^{i} (i=1,\cdots,N)$.\\

\textbf{Step~2.} For $i=1,\cdots,N$, $CI^{i}$ is divided into $\eta^2$ parts
as $CI^{i}_{j}~(i=1,\cdots,N,~j=1,\cdots,\eta^2)$
, where 
\begin{align}
\eta:=
\left\{%
\begin{array}{ll}
2,&~when~N~is~1,~2 ~or~3,\\
\\
\lfloor \sqrt{N} \rfloor,&~otherwise.\\
\end{array}%
\right.
\end{align}
Then we consider a  block matrix as $\Theta=(CI^i_{j})$ (see figure \ref{f-2}).\\

\textbf{Step~3.}  \\

a) $v^l~(l=1,2,3)$ are defined as $v^l=\mu(x^l_0,r_l,N\eta^2)$ with $x^l_0:=\frac{\sum_{i,j}{CI^{1}(:,:,l)}}{nc\times mc\times 255}~(l=1,2,3)$.\\

b)  $Sv^l:=sort(v^l)~(l=1,2,3)$  (smallest to largest).\\

c) $A^l$ and $P^l$ are  considered as $A^l=\Delta(V^l),~P^l=\Delta(SV^l)~(l=1,2,3)$.\\

d) $C$ is defined as  $C(:,:,l):=\Upsilon^2_{A^l\rightarrow P^l}(\Theta(:,:,l))~(l=1,2,3).$\\ 

\textbf{Step~4.}\\

a) To generate different keys in each iteration, we consider $x^0$
as random number in $(0,1)$ then we define $v^0$ as
$\mu(x^0_0,r_0,N\times nc\times mc)$.\\

b)  $Sv^0:=sort(v^0)$  (smallest to largest).\\

c) Consider $A^0=\Lambda(v^0),~P^0=\Lambda(Sv^0)$.\\

d) By using $xor$ operation and $\Upsilon^3_{A^0\rightarrow P^0}$, matrix $EI$ is  determined by
\begin{align*}
EI:=\Upsilon^3_{A^0\rightarrow P^0}(C)\oplus A.
\end{align*}

e) The final encrypted image is found as follows
\begin{align*}
EI=circshift(EI,[\rho_1~\rho_2~\rho_2]),
\end{align*}
where $circshift(A,r)$ circularly shifts the values in array $A$ by $r$ positions and
$\rho_l=\lfloor x_l\times10^2\rfloor,~l=1,2,3$.

\subsection{Decryption algorithm}
The decryption algorithm is the inverse process of the encryption algorithm.
In the first step by using Algorithm \ref{AL1}, compressed images 
are found. Then by using inverse of the F-transform we can find images.
\begin{algorithm}\label{AL1}
\SetKwInOut{Input}{Input}
\SetKwInOut{Output}{Output}
\Input{$EI_{N\times nc\times mc}$ (Compression-encryption image),\\
 $\{x_i\}^{3}_{i=0},~\{r_i\}^{3}_{i=0}$ (Keys),}
\Output{$\{CI^i\}^{N}_{i=1}$ (Compressed images),}
\For{l=1:3}{
$V^l:=\mu(x_l,r_l,N\eta^2);$\\
$A^l:=\Delta(V^l), P^{l}:=\Delta(sort(V^l));$\\
$\rho_l=\lfloor x_l\times10^2\rfloor;$\\}
$V^0:=\mu(x_0,r_0,N\times nc\times mc);$\\
$A^0:=\Lambda(V^0), P^{0}:=\Lambda(sort(V^0));$\\
$EI=circshift(EI,[\rho_1~\rho_2~\rho_3]);$\\
$C=\Upsilon^3_{P^0\rightarrow A^0}(EI)\oplus A;$\\
\For{l=1:3}{
$\Theta(:,:,l):=\Upsilon^2_{P^l\rightarrow A^l}(C(:,:,l));$
}
\caption{Decryption Algorithm}
\end{algorithm}
\section{Experimental results and security analysis}
\subsection{Experimental results}
In this section, as plain images, we use color images  ``Baboon,
Lena, Peppers and House'' (the $256\times 256$ images with 256 grey levels)
as $I^1,I^2,I^3$ and $I^4$ respectively.
In compression step $0.1$ is used as tension parameter.
Also we consider $nc=mc=100$.
Fig. 4 shows original images and compressed images by using 
exponential b-spline
fuzzy transform.
Also simulation results for encryption algorithm are shown 
in Fig. 5. In this section, we consider the secret keys as
$r_0=3.02,r_1=r_2=r_3=2$. Also for simulation results, 
we use Matlab R2014b programming.
\begin{center}
\begin{figure}
\centering
\includegraphics[width=140mm,scale=0.70]{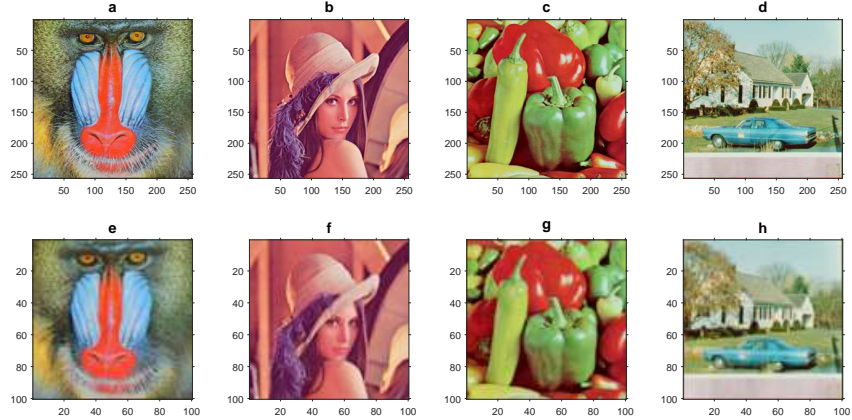}
\emph{\caption{(a-b-c-d) Original images, (e-f-g-h) Compressed images.}}
\end{figure}
\end{center}  
\begin{center}
\begin{figure}
\centering
\includegraphics[width=140mm,scale=0.70]{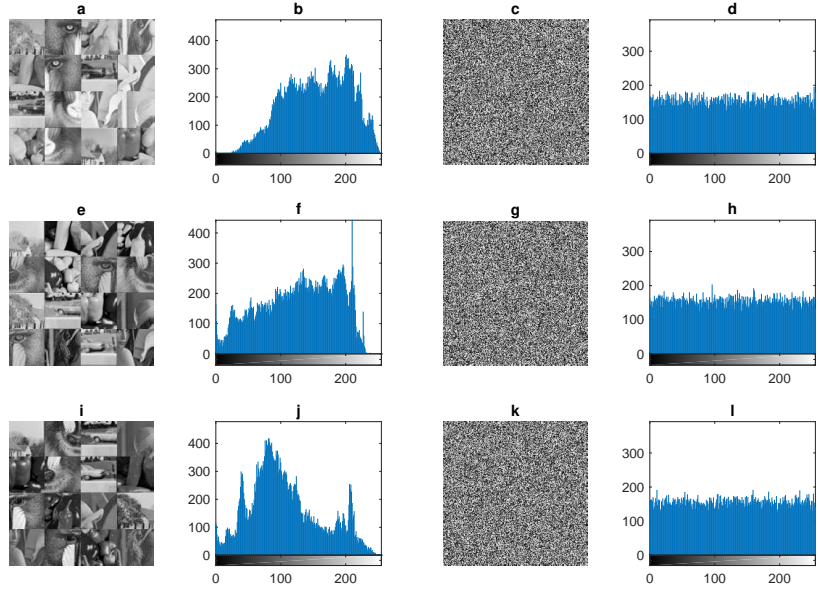}\label{f-3}
\emph{\caption{(a-e-i) R, G, B components of the step 2 process of the proposed algorithm,
(b-f-j) Histograms of  R, G, B components of the step 2 process of the proposed algorithm,
 (c-g-k), R, G, B components of the encrypted
image, (d-h-l) Histograms of  R, G, B components of the encrypted
image.
}}
\end{figure}
\end{center} 
\subsection{Key analysis}
In the encryption algorithm, the security keys of the proposed algorithm
are composed of eight parameters. 
In the image
encryption algorithm, if we use $10^{15}$ as the precision,
 the key space is almost
$10^{120}$, and this space is sufficiently large to resist the brute force attack \cite{16}.
In the next step, we study key sensitivity. We change $r_0$ as $r_0=3.02+10^{-15}$.
Fig. 6. shows decryption images by using changed key.
By using this figure, we can see that the original images cannot be reconstructed. 
Then, we can see that the proposed
algorithm has high key sensitivity. Also in Step 4 of encryption algorithm, a random number is used as encryption key. 
Therefore we can  generate different keys in each iteration.
We run the encryption algorithm twice and
by using pixel-to-pixel the difference
between the two images are illustrate in Fig. 7.
It can
be seen in Fig. 7, two encrypted images are different. Therefore the proposed algorithm
is able to resist the chosen-plain text attack.

 \begin{center}
\begin{figure}
\centering
\includegraphics[width=110mm,scale=0.5]{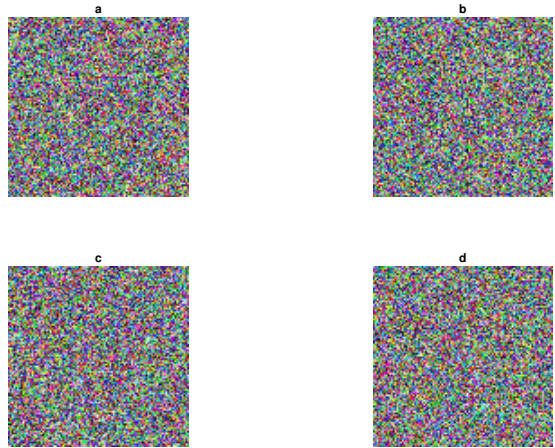}
\emph{\caption{Decrypted images by using 
changed key.}}
\end{figure}
\end{center} 

 \begin{center}
\begin{figure}
\centering
\includegraphics[width=110mm,scale=0.5]{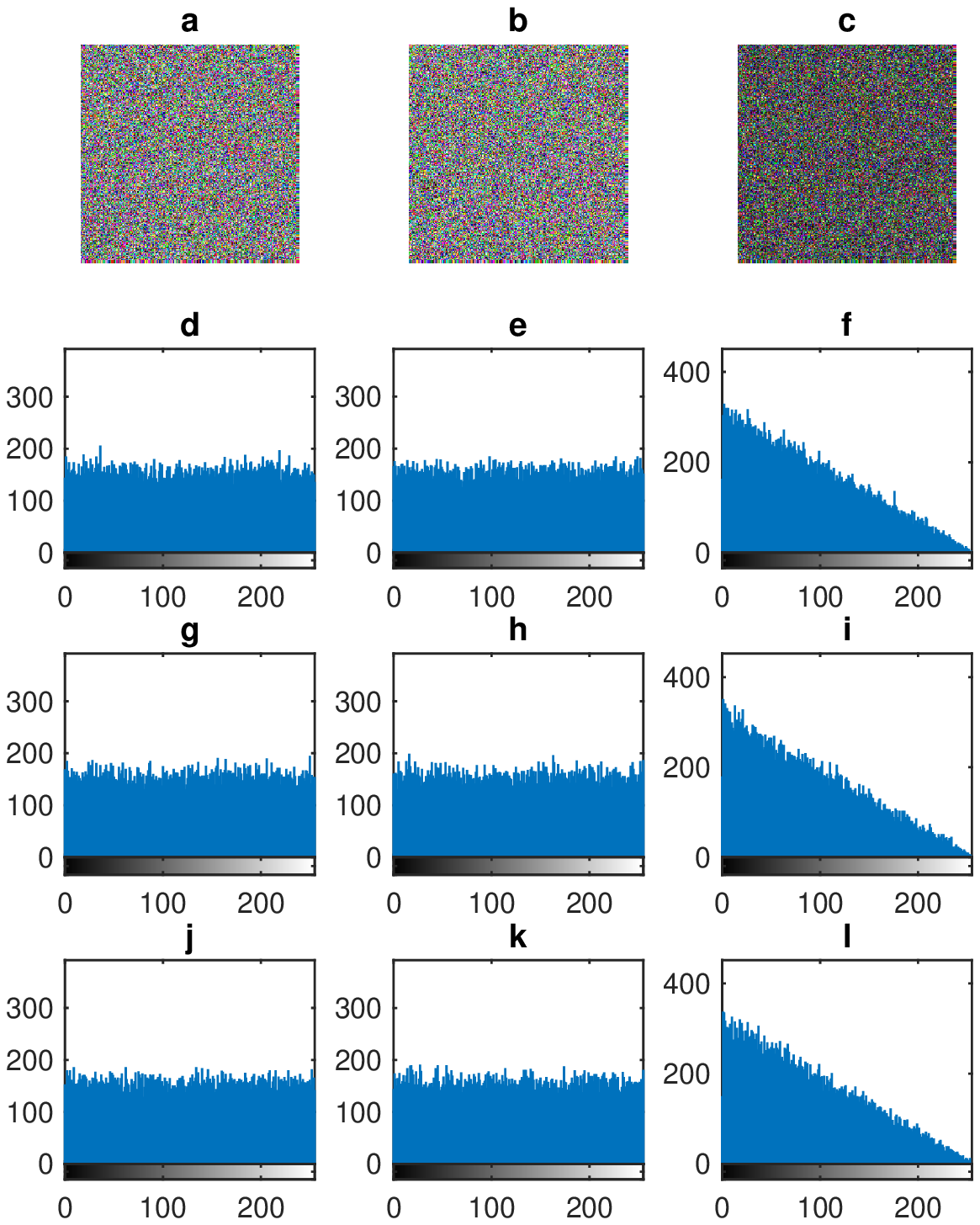}
\emph{\caption{
(a-d-g-j) The first encrypted image and
 histogram of  R, G, B components , (b-e-h-k) The second encrypted image and
 histogram of  R, G, B components, (c-f-i-l) The pixel-to-pixel
difference and  histogram of  R, G, B components.}}
\end{figure}
\end{center} 

\subsection{Noise and Data loss attacks}
Appropriate encryption algorithm 
should resist the
data loss and noise attacks.
In the data loss attack, some image data
disappears. To simulate this attack, in Fig. 8,
we remove $100\times100$ of encrypted
image. Also in the noise attack, noise is added to the encrypted image.
An appropriate encryption algorithm should not increase the amount of these noises.
To simulate noise attack, Gaussian noise with zero-mean and $var=0.2$ is added to
encrypted image. The simulation results are given in Fig.s 8-9. 
In  these figures we can see that
 the reconstructed images contain
most of original visual in formation and we can recognize the original images.
\begin{center}
\begin{figure}
\centering
\includegraphics[width=110mm,scale=0.5]{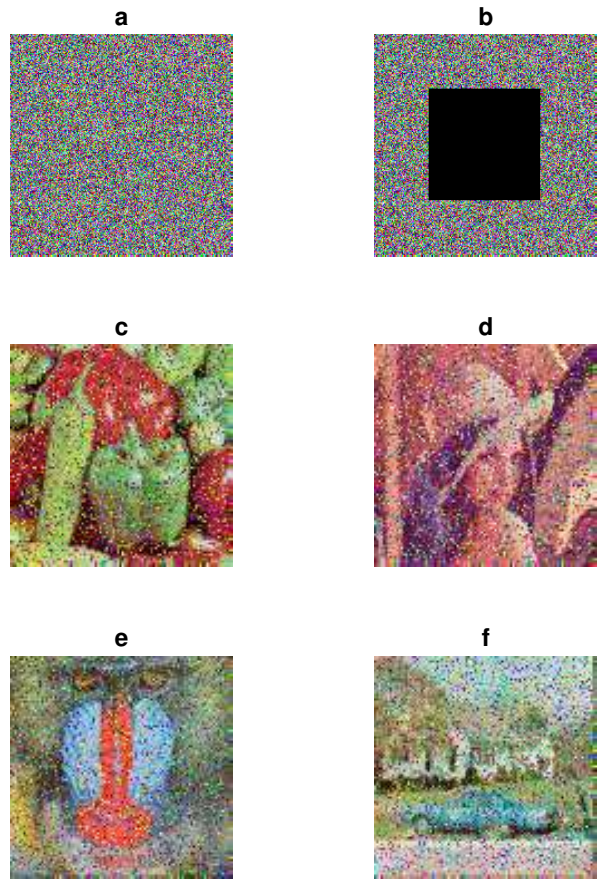}
\emph{\caption{(a) Encrypted image, (b) Cropped attack image, (c-d-e-f)
Decryption results.}}
\end{figure}
\end{center}  
\begin{center}
\begin{figure}
\centering
\includegraphics[width=110mm,scale=0.5]{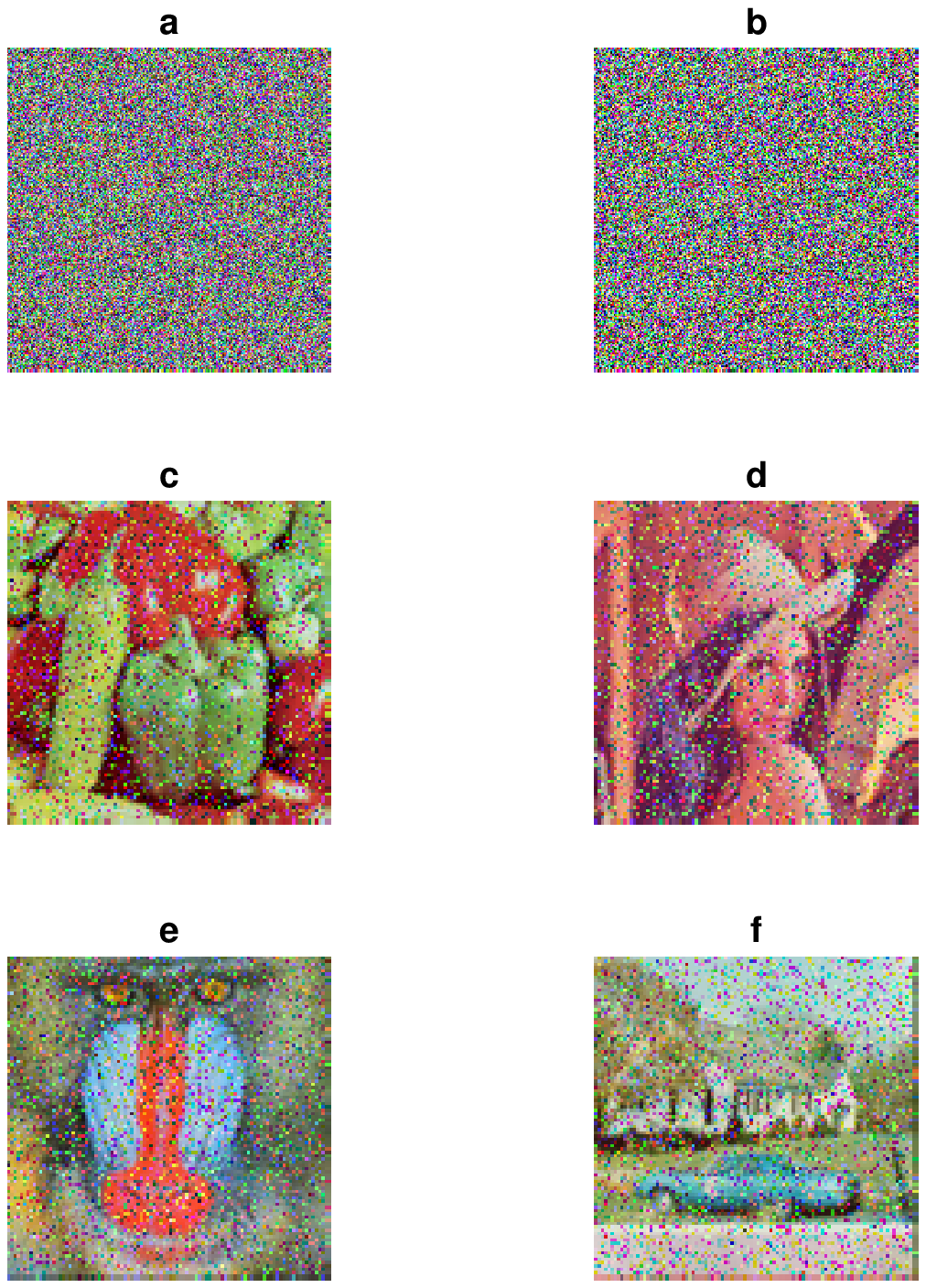}
\emph{\caption{(a) Encrypted image, (b) Noise attack image with
var=0.2, (c-d-e-f)
Decryption results.}}
\end{figure}
\end{center}  
\subsection{Statistical analysis}
Some statistical analysis as  correlation values, information entropy are studied in 
this section. In the appropriate encryption algorithm, the correlation values 
between adjacent pixels are close to zero. 
The correlation coefficients of adjacent pixels in the plain images
and  encrypted images are given in Table 1 and Fig.s 10-11.
By using numerical results, we can say that
the correlation coefficients of plain images are close to one, and the
correlation coefficients of the encrypted image are close to
zero.
The correlation values are found by using formula
 \cite{17}
\begin{align}
C_{xy}=\frac{E[(x-\mu_x)(y-\mu_y)]}{\sigma_x \sigma_y},
\end{align}
where $E[\cdot]$ denotes the expectation value, $\mu$ is
the mean value and $\sigma$ represents standard deviation.

In the next step we study histogram analysis.
The evaluation of the robustness
of an encrypted image is studied by using histogram analysis.
The histograms of  ciphered image are shown in Fig. 5.
The histograms show that
distribution of pixel values in the
encrypted image is close to uniform distribution.
Also chi-square test outputs are given in Table 2.
By using chi-square test, we accept that the data 
have a uniform distribution at level 0.05.
Also unpredictability of information can be studied by using unpredictability of information
the information entropy.  The information entropy is defined as follows \cite{18}
\begin{align}
H(k)=-\sum^{w-1}_{i=0}P(k_i)\log_2P(k_i),
\end{align}
where $w$ is the gray level and $P(\cdot)$ denotes the probability of symbol.
The ideal value of the information entropy is 8. Numerical results for the information entropy 
are given in Table 3. 
From Table 3, we can say
that  results for the encryption algorithm are very close to the ideal value.
From the above discussion, we can conclude that the proposed algorithm
has stronger ability to withstand statistical attacks.
\begin{center}
\begin{table}[ht!]
\caption{{\footnotesize Correlation coefficient values in the plain images and the cipher image.}}
\label{table:t1}
\centering
\begin{tabular}{cccccc}
\hline
Image            &Component        &Horizontal      &Vertical  &Diagonal            &Diagonal \\
                     &                         &                    &            &\tiny{ (lower left to top right)}&\tiny{(lower right to top left)}\\
\hline \hline
                      &R                       &0.9635      &0.9663   &0.9585      &0.9564     \\
Baboon           &G                       &0.9811      &0.9818    &0.9709      & 0.9687    \\
                      &B                       &0.9665      &0.9664    &0.9478      &0.9478   \\
\hline
                      &R                      &0.9798     &0.9893   &0.9777       &0.9697     \\
Lena               &G                      &0.9691      &0.9825   &0.9654       & 0.9555    \\
                      &B                      &0.9327     &0.9576    &0.9254     &0.9183    \\
\hline
                     &R                       &0.9231     &0.8660   &0.8519       &0.8543   \\
Peppers          &G                      &0.8655     &0.7650    &0.7249       &0.7348    \\
                     &B                       &0.9073      &0.8809   &0.8424      &0.8399   \\

\hline
                   &R                       &0.9536      &0.9579   &0.9240       &0.9224   \\
House          &G                       &0.9391     &0.9423   &0.8940      &0.8901     \\
                  &B                        &0.9725     &0.9686   &0.9459       &0.9445\\
\hline
                     &R                     &0.0070      &0.0025    &-0.0066      &-0.0014     \\
Ciphered image  &G                     & 0.0019      &0.0072    &0.0030       &0.0056    \\
                    &B                      &-0.0004      &0.0019    &0.0048       &0.0008\\
\hline
\end{tabular}
\end{table}
\end{center}
\begin{center}
\begin{table}[ht!]
\caption{{\footnotesize  Chi-square outputs for ciphered image.}}
\label{table:t1}
\centering
\begin{tabular}{cccccc}
\hline
                               & R Component       & G Component      &B Component  \\
\hline
\hline
Chi-square statistics    &236.6           &288.1            &240.5        \\
df                              &255                 &255                    &255          \\
p-value                      &0.790               &0.076                 &0.735       \\
\hline
\end{tabular}
\end{table}
\end{center}
\begin{center}
\begin{table}[ht!]
\caption{{\footnotesize  Information entropy,  NPCR and UACI results.}}
\label{table:t1}
\centering
\begin{tabular}{cccccc}
\hline
            & Information entropy        & NPCR     &UACI  \\
\hline
\hline
R Component         &7.9955     &99.6250    &33.4561         \\
G Component         &7.9945      &99.6400    &33.4212          \\
B Component         &7.9952     &99.6250  &33.3425       \\
\hline
\end{tabular}
\end{table}
\end{center}
\subsection{Sensitivity analysis}
NPCR (number of pixels change rate) and UACI (unified average change intensity)
are calculated as
\begin{align}\label{ex1}
&NPCR=\frac{\sum_{i,j}D(i,j)}{m\times n}\times 100\%,\\
&UACI=\frac{1}{m\times n}\big [ \frac{\sum_{i,J}|C_1(i,j)-C_2(i,j)|}{255}\big]\times 100\%,
\end{align}
where
\begin{align}\label{ex1}
&D(i,j):=\left\{%
\begin{array}{ll}
1,&when~C_1(i,j)\neq C_2(i,j),\\
\\
0,&when~C_1(i,j)= C_2(i,j).\\
\end{array}%
\right.
\end{align}
In above formulae, $C_1(i,j)$ and $C_2(i,j)$ are denoted  the cipher
image before and after one pixel of the plain image
is changed. 
The sensitive to the changing of plain image can be studied
using NPCR and UACI. The best value for NPCR is $100\%$.
Also the ideal value for UACI is $33.\overline{33}\%$.
In this step,  $I_1(10,50,2), I_2(20,30,3), I_3(60,35,1)$  and $I_4(75,34,2)$ are changed
to 0.
The simulation results in Table 3 
are close to ideal values.
Therefore, we can say that the differential attack is impossible on
proposed algorithm.

\begin{center}
\begin{figure}
\centering
\includegraphics[width=110mm,scale=0.5]{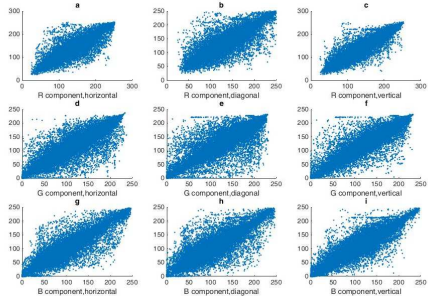}
\emph{\caption{ Correlation of neighbourhood pixels at different directions of $\Theta$. }}
\end{figure}
\end{center}  
\begin{center}
\begin{figure}
\centering
\includegraphics[width=110mm,scale=0.5]{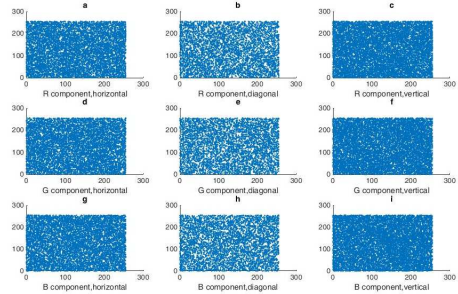}
\emph{\caption{Correlation of neighbourhood pixels at different directions of the encrypted image. }}
\end{figure}
\end{center}  
  
\section{Conclusion}
In this paper, by using fuzzy transform and 
combination chaotic system,
we have introduced a 
multi-color image
compression-encryption algorithm.
In fuzzy transform, exponential b-spline function is used as
fuzzy partition.
Simulation results shown that 
the proposed encryption algorithm can effectively resist differential, statistical, noise, data loss, chosen-plain text attacks.


\begin{thebibliography}{99}
\providecommand{\doi}[1]{DOI~\discretionary{}{}{}#1}
\bibitem{1}
A. Kanso, M. Ghebleh. {\em An algorithm for encryption of secret images into meaningful images}. Optics and Lasers in Engineering 90 (2017): 196-208.
\bibitem{2}
N. Zhou, A. Zhang, F. Zheng, L. Gong. {\em Novel image compression–encryption hybrid algorithm based on key-controlled measurement matrix in compressive sensing}. Optics $\&$ Laser Technology 62 (2014): 152-160.
\bibitem{3}
Z. Li, X. Sun, Q. Ding. {\em Design and implementation in image compression encryption of digital chaos based on matlab}. In Intelligent Data analysis and its Applications, Volume II, pp. 509-518. Springer, Cham, 2014.
\bibitem{4}
N. Jindal,  K. Singh. {\em Joint image compression–encryption using discrete fractional transforms}. The Imaging Science Journal 62, no. 5 (2014): 265-272.
\bibitem{5}
Zhang, Liguo, Binghang He, Jianguo Sun, Mingzhu Lai, and Zhihan Lv. "Double image multi-encryption algorithm based on fractional chaotic time series." Journal of Computational and Theoretical Nanoscience 12, no. 11 (2015): 4980-4986.
\bibitem{6}
{\"U}. {\c C}avu{\c s}o{\u g}lu, S. Ka{\c c}arb, I. Pehlivanb, A. Zengina, {\em Secure image encryption algorithm design using a novel chaos based S-Box},
Chaos, Solitons $\&$ Fractals 95 (2017) 92-101.
\bibitem{7}
D. Kong, X. Shen, Y. Shen,  Xin Wang. {\em Multi-image encryption based on interference of computer generated hologram}. Optik-International Journal for Light and Electron Optics 125, no. 10 (2014): 2365-2368.
\bibitem{8}
S. Banerjee, S. Mukhopadhyay,  L. Rondoni. {\em Multi-image encryption based on synchronization of chaotic lasers and iris authentication}. Optics and Lasers in Engineering 50.7 (2012): 950-957.
\bibitem{9}
I. Perfilieva. {\em Fuzzy transforms: Theory and applications}. Fuzzy sets and systems 157, no. 8 (2006): 993-1023.
\bibitem{10}
F.D. Martino, V. Loia, I. Perfilieva, S. Sessa. {\em An image coding/decoding method based on direct and inverse fuzzy transforms}. International Journal of Approximate Reasoning 48, no. 1 (2008): 110-131.
\bibitem{11}
I. Perfilieva, V. Nov{\'e}k, A. Dvo{\v{r}}{\'a}k. {\em Fuzzy transform in the analysis of data}.International Journal of Approximate Reasoning 48, no. 1 (2008): 36-46.
\bibitem{12}
P.E. Koch, T. Lyche. {\em Interpolation with exponential B-splines in tension}. In Geometric modelling, pp. 173-190. Springer, Vienna, 1993.
\bibitem{14}
I. Bero{\v{s}}, M. Maru{\v{s}}i{\'c}. {\em Evaluation of tension splines}. Mathematical Communications 4, no. 1 (1999): 73-81..
\bibitem{15}
R. Parvaz, M. Zarebnia. {\em A combination chaotic system and
application in color image encryption}. 	arXiv:1708.01862.
\bibitem{16}
G. Alvarez, S. Li. {\em Some basic cryptographic requirements for chaos-based cryptosystems}. Int J Bifurcation Chaos 2006;16(8):2129–51.
\bibitem{17}
N. Zheng, J. Xue. {\em Statistical learning and pattern analysis for image and video processing}.
Springer Science $\&$ Business Media, 2009.
\bibitem{18}
F.A El-Samie, H.H. Ahmed, I.F. Elashry, M.H. Shahieen, O.S. Faragallah,
E.M. El-Rabaie, S.A. Alshebeili. {\em Image Encryption: A Communication
Perspective}. CRC Press 2014.

\end{thebibliography}
\end{document}